\documentstyle[epsf]{elsart}
\newcommand{\bea}{\begin{eqnarray}}
\newcommand{\eea}{\end{eqnarray}}
\newcommand{\be}{\begin{equation}}
\newcommand{\ee}{\end{equation}}

\begin{document}
\begin{frontmatter}
\title{Liquid Crystals in the Mantles of Neutron Stars}
\author {C. J. Pethick$^{1,2}$ and A. Y. Potekhin$^{1,3}$}
\address{$^{1}$ NORDITA, Blegdamsvej 17, 
DK-2100 Copenhagen \O, Denmark \\
$^{2}$ Department of Physics, University of Illinois 
at Urbana-Champaign,\\
1110 West Green St., Urbana, Illinois 61801-3080, USA\\
$^3$ Ioffe Physical-Technical Institute,
           Politekhnicheskaya 26,
           194021 St. Petersburg, 
           Russia
}


\begin{abstract}
Recent calculations indicate that in the outer parts 
of neutron stars nuclei 
are rod-like or slab-like, rather than roughly spherical.  
We consider the elastic properties of these phases,
and argue that they behave as liquid crystals,
rather than rigid solids. We estimate elastic constants
and discuss implications of our results for neutron star behavior.
\end{abstract}

\begin{keyword}
Dense matter.
Liquid crystals.
Elasticity, elastic constants.
Neutron stars.\\
PACS numbers: 26.60.+c, 61.30.-v, 62.20.Dc, 97.60.Jd
\end{keyword}

\end{frontmatter}


In the standard picture~\cite{shapiro}, the outer part of a neutron star is 
a solid from the bottom of the ocean of melted iron 
at a density $\rho\sim10^6-10^8$ g cm$^{-3}$ (depending on temperature)
down to the boundary with the core of the star at a density of order the
saturation density of nuclear matter,
${\rho}_{s} \sim 3 \times10^{14}$ g cm$^{-3}$,
and matter consists of roughly spherical atomic nuclei 
arranged on a regular bcc lattice.  
Charge neutrality is provided by a background of electrons, 
and, at densities in excess of about $4\times 10^{11}$ g cm$^{-3}$, 
there is a neutron liquid between the nuclei.
However, recent work indicates that, in roughly half 
of the matter by mass 
in what is traditionally referred to as the ``crust" of a neutron star, 
the nuclei are very aspherical, 
and have the form of essentially infinitely long rods or infinitely 
extended slabs (for a review, see Ref.~\cite{pr95}).
Following Ref.~\cite{dG}, we shall refer to these phases 
as mesomorphic phases, since they have properties intermediate 
between liquids and solids.
  
The possibility of nuclei in matter at sub-nuclear densities 
adopting very non-spherical shapes was first pointed out 
in the context of stellar collapse
by Ravenhall et al.~\cite{rpw}, and confirmed 
by other authors using other models~\cite{hashimoto,ohy,wk85,lassaut}.
With increasing density, nuclei become first rods, then slabs. 
This is followed by two ``inside-out" phases in which  there are 
cylindrical or spherical ``bubbles" in the nuclear matter.  
Subsequently the transition to the uniform liquid phase occurs.  
For neutron star matter with dripped neutrons,
Lorenz et al.~\cite{lrp} found a similar sequence of phase transitions, 
and these conclusions were confirmed by other 
investigators~\cite{oyamatsu,oy}.

Mechanical properties of the outer parts of neutron stars are important in 
attempting to understand many aspects of neutron star behavior,
including  glitches in the pulse repetition rate, magnetic field evolution, 
and neutron star models for $\gamma$-ray bursts~\cite{alpar,ruderman,blaes}.
Our purpose in this Letter is to discuss the elastic properties 
of the mesomorphic phases and estimate elastic constants.   
 
We shall concentrate on the phases with
rod-like and slab-like nuclei, since they are expected to be
the dominant ones.
According to Ref.~\cite{lrp}, these 
``spaghetti" and ``lasagna" phases constitute, respectively, 
roughly $\frac{1}{5}$ and $\frac{1}{2}$ of the mass lying
between the core and the ordinary solid crust, 
where nuclei are roughly spherical. 
In Ref.~\cite{oy} these proportions
are found to be about $\frac{1}{2}$ and $\frac{1}{3}$. 
Thus the fraction of mass in the bubble phases is relatively 
small. 

As a model, we shall assume uniform rods and slabs. 
For symmetric nuclear matter, Thomas--Fermi
calculations~\cite{wk85}
have confirmed that these simple configurations are 
the thermodynamically favorable ones.  
Clearly there is no increase in energy if rods or slabs 
are displaced in directions that lie in the plane of the slabs, 
or along the rods. 
Consequently there is no restoring force for certain sorts of distortion, 
and they thus have the elastic properties of liquid crystals. 
In the ``lasagna" and ``spaghetti" phases,
microscopic calculations indicate that there is positional order 
in one and two directions,
respectively, maintained by
     the Coulomb repulsion of rods and
     slabs~\cite{rpw,hashimoto,ohy,wk85,lassaut,lrp}.
 Accordingly, they conform to the  
definitions of  columnar phases and 
     smectics A~\cite{dG}.
     More complex positionally ordered phases (e.g., smectics C,
     cholesterics) are precluded by the symmetry of the equilibrium
     shapes of the nuclei.  At the temperatures of
neutron star interiors ($\sim 10^8$~K) positionally 
disordered (nematic) phases
are
unlikely, since one would expect the ordering temperature of rods and plates
to be comparable with the melting temperature for matter with spherical
nuclei, $\sim 10^{10}$~K.
We emphasize, however, that the physical reasons 
for the spatial structure of the mesomorphic phases 
in the laboratory and in neutron stars are very different.  
For laboratory liquid crystals, the non-spherical shape 
of the molecules drives the 
tendency to form rod-like and slab-like structures, 
while in neutron stars, it is a spontaneous symmetry 
breaking brought about by the competition between 
the nuclear surface energy and the Coulomb energy.
In the neutron star case, the basic objects, nucleons, from which 
structures are constructed are spherical, while in laboratory liquid 
crystals, the basic ingredients are non-spherical molecules. 

To calculate energies of these phases we adopt a generalized 
liquid drop model, with bulk, surface and Coulomb energies.  
In the deformations that we study in this paper we assume that
the total density remains constant. Distortion can lead to 
a redistribution of neutrons between nuclear matter and neutron matter, 
but this is  
small because bulk energy densities 
are large compared with those of surface and Coulomb energies.  
Therefore we may assume that the fractions of the total volume occupied 
by nuclear matter and neutron matter, and their local densities remain 
constant, and consequently only the Coulomb 
and surface energies are altered. 

\begin{figure}[h]
\begin{center}
\leavevmode
\epsfysize=37mm
\epsfbox[170 380 470 530]{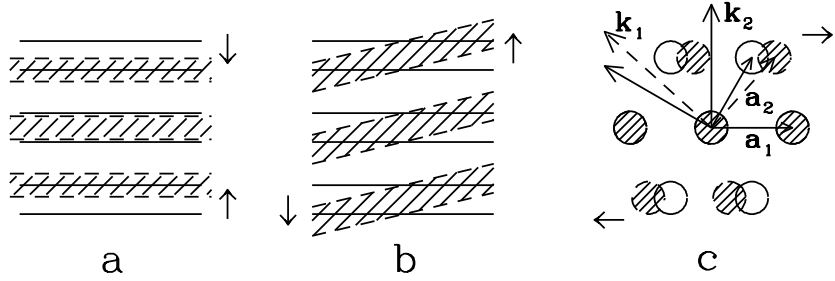}
\end{center}
\caption{Linear deformations.
Cross sections of the slabs and rods in equilibrium are indicated by
full lines, and, after deformation,
by the hatched areas. 
a: compression perpendicular to the slabs (rods);
b: shear;
c: transverse shear in the columnar phase;
the primitive translation vectors
(${\bf a}_1$, ${\bf a}_2$) and their
counterparts in the reciprocal lattice
(${\bf k}_1$, ${\bf k}_2$)
are also shown.
}
\label{fig1}
\end{figure}

We begin by estimating the elastic constants for the 
layered phase.  
Since there is complete rotational symmetry 
  about the axis perpendicular to the layers, which we denote by $Oz$, 
  this phase is similar to a smectic A liquid crystal,
and the energy density due to deformation
may be written in the form~\cite{dG}
\be
  E_d = {B\over 2}\left[
  {\partial u\over\partial z} - {1\over2}(\nabla_\perp u)^2
  \right]^2
  + {K_1\over2}(\nabla_\perp^2 u)^2,
\label{Ed1}
\ee
where $u$ is the displacement of layers in the $z$ direction.
The first term is associated with a change of interlayer distance
(Fig.~\ref{fig1}a). The shear shown in Fig.~\ref{fig1}b is equivalent, 
due to rotational invariance, to reducing the interlayer
spacing, an effect taken into account by the second term 
in the square brackets. 
The second-order elastic constant $K_1$
is associated with splay deformations.
As a first case, we consider a distortion that maintains the planar 
character of the slabs but changes the layer spacing, 
which we denote by $2r_c$.
The surface energy per unit volume, $E_\sigma$
scales as $r_c^{-1}$, while the Coulomb energy per unit volume, 
$E_C$ scales as $r_c^2$, 
and therefore in equilibrium these energies satisfy the condition
$E_{\sigma0}=2E_{C0}$~\cite{pr95,rpw}, which determines $r_c$.  
Here the subscript ``0" denotes 
equilibrium values. From the same scaling law we conclude
that a small departure from equilibrium 
associated with a change $\delta r_c$ 
of $r_c$ results in an increase of energy
\be
  \delta E_{\sigma+C} = 3E_{C0} (\delta r_c/r_c)^2.
\label{dE}
\ee
Applying Eqs.~(\ref{Ed1})   
to the same deformation, we obtain $E_d=\frac{1}{2}B(\delta r_c/r_c)^2$, 
whence it follows that
\be
  B=6E_{C0}.
\ee

In order to estimate $K_1$, let us consider a harmonic perturbation
$u(x)=u_0\cos(kx)$, assuming that its wavelength is
large: $(kr_c)^2\ll1$ and $(ku_0)^2\ll1$. Then the deformation energy
averaged over $x$ is
\be
  \langle E_d \rangle = \frac{3}{64}B\,(ku_0)^4 + 
  \frac{1}{4}K_1 k^4 u_0^2.
\label{elastic}
\ee
The first term in this expression dominates 
when $u_0\gg\lambda$, where $\lambda=(K_1/B)^{1/2}$.
To order $k^4$,
the relative change in the surface energy 
averaged over $x$ is
\be
  {\langle \Delta E_\sigma \rangle \over E_{\sigma0}} = 
  {1\over4}\,(ku_0)^2 - {3\over64}\,(ku_0)^4.
\label{surface}
\ee
If one includes curvature contributions to the surface energy one finds 
that the first order terms average to zero, while the second order ones 
give an additional contribution 
$b k^4u_0^2/2$ to this equation, where $b$ is the ratio of the second-order
curvature coefficient to the surface tension.  For
symmetric nuclear matter, according to the estimate of 
Bennett and Ravenhall~\cite{br74}, 
$b\approx -(0.4{\rm\ fm})^2$.

It is convenient to express the charge density of nuclei as a Fourier series, 
and for the distortion given above one finds
\be
  \rho(x,z)=\sum_{mn}\rho_{mn}\exp({\rm i}m\pi z/r_c+{\rm i}nkx).
\ee
Here 
$\rho_{mn}=\rho_p\,(m\pi)^{-1}\sin(m\pi r_N/r_c)
(-{\rm i})^n J_n(m\pi u_0/r_c)$,
$\rho_p$ being the charge density of protons inside nuclei,
$r_N$ the halfwidth of the slab,
and $J_n$ the Bessel function of order $n$,
and the summation is over all integers.
The spatial average of the Coulomb energy is
\be
  \langle E_C\rangle = {\sum_{mn}}' {2\pi\, |\rho_{mn}|^2\over
  k_{TF}^2 + (m\pi/r_c)^2 + (nk)^2 },
\label{EC1}
\ee
where $k_{TF}$ is the Thomas--Fermi screening wavenumber,
and the prime on the summation indicates that
the term with $m=n=0$ is excluded. 
In the absence of distortion ($u_0=0$), the screening correction can be 
neglected in  evaluating the Coulomb energy, 
since $k_{TF}^2\ll r_c^{-2}$, as argued in Refs.~\cite{wk85,lprl}, 
and the summation over $m$ in Eq.~(\ref{EC1}) yields
$E_{C0}=2\pi\,(\rho_p r_N)^2 (1-w)^2/3$, in agreement with Ref.~\cite{rpw}.
Here $w$ is the fraction of space occupied by nuclear matter, and $w=r_N/r_c$
for the smectic case.  From the equilibrium condition $E_{\sigma 0}= 2E_{C
0}$ and the fact that the surface energy is given by $E_{\sigma 0}=w
\sigma/r_N$, where $\sigma$ is the surface tension of the interface between
nuclear matter and neutron matter, it follows that in equilibrium
$r_N=(3w\sigma)^{1/3}(4\pi\rho_p^2 (1-w)^2)^{-1/3}$.  
Thus the Coulomb energy in
equilibrium varies as the two thirds power of the surface tension.

Although $k_{TF}$ may be
large compared with $k$, the effects of screening may be neglected 
also in 
calculating the change in the Coulomb energy due to distortion,
$\langle \Delta E_C\rangle = \langle E_C\rangle - E_{C0}$, 
because $\rho_{0n}=0$ for any nonzero value of $n$.  Thus we find 
\be
  \langle \Delta E_C\rangle = (\rho_p\,r_N)^2
  \sum_{j\geq1} (-1)^j a_j(\xi)\,(kr_c/\pi)^{2j},
\label{Coulomb}
\ee
where
\be
  a_j(\xi) = {8\over\pi^3 w^2} \sum_{m=1}^\infty 
  {\sin^2(m\pi w)\over m^{4+2j}}
  \sum_{n=1}^\infty n^{2j}\,J_n^2(m\xi),
\ee
with $\xi = \pi u_0/r_c$. In particular, 
we obtain $a_1(\xi)=\pi \xi^2 \,(1-w)^2/3$ and
$
  a_2(\xi) =\pi \xi^2 \, (1-w)^2 \,
  [ (1+2w-2w^2)\,\pi^2 /45 + \xi^2/4 ].
$
The leading Coulomb contribution to the energy 
due to distortion is $ -E_{C0} (ku_0)^2/2$.  
This exactly cancels the corresponding surface term 
in Eq.~(\ref{surface}), 
as it must on general grounds.
By comparing the sum of the Coulomb (Eq.~(\ref{Coulomb})) 
and surface (Eq.~(\ref{surface})) 
contributions to the energy due to  distortion with the general 
expression (\ref{elastic}) we find
 \be
  K_1 = 2\,E_{C0}\,(1+2w-2w^2)\,r_c^2/15.
\ee
The curvature correction to this expression is $4bE_{C0}$, 
which may be neglected since  $r_c$ is typically $\sim10$ fm. 

In the calculations above we assumed that the thickness of a slab
{\it measured along $Oz$\/} remains unperturbed. In reality,
the slab will be thinner near the extrema of $u(x)$
and thicker near the extrema of $\partial u/\partial x$. 
Allowance for this effect, however, 
yields a contribution to $\langle E_d\rangle$ containing an additional
small factor $\,\sim(k^2+k_{TF}^2)\,r_c^2$ relative to the terms
already considered. 

Finally, we remark that, even though at finite temperatures strict long range 
positional order of layered phases will be destroyed according to the standard 
Landau-Peierls argument, our theoretical estimates of elastic constants should 
be a good approximation at low temperatures.

Now we turn to the columnar phase. 
We assume the rods to have a circular cross section of radius $r_N$
and define $r_c$ 
by the condition that the density of rods per unit area is $1/\pi r_c^2$. 
Then the scaling laws that lead to Eq.~(\ref{dE}) hold also for this case.

The lowest energy configuration 
is one with rods on a two-dimensional triangular lattice.  We choose $Oz$
to be the axis of the D$_{\rm h}$ symmetry.
Displacements are then described by a two-dimensional vector
${\bf u}=(u_x,u_y)$, and the energy of deformation  
is
\bea
  E_d &=& {B\over2}\left(
  {\partial u_x\over\partial x} +
  {\partial u_y\over\partial y}\right)^2
  + {C\over2}\left[ 
  \left({\partial u_x\over\partial x} -
  {\partial u_y\over\partial y}\right)^2
\right.
\nonumber\\
  &+& 
\left.
  \left({\partial u_x\over\partial y} +
  {\partial u_y\over\partial x}\right)^2 
\right]
  +{K_3\over2}\left({\partial^2 {\bf u}\over\partial z^2}\right)^2
\nonumber\\
  &+&
  B'\left({\partial u_x\over\partial x} +
  {\partial u_y\over\partial y}\right)
  \,\left({\partial{\bf u}\over\partial z}\right)^2
  + {B''\over2}\left({\partial{\bf u}\over\partial z}\right)^4.
\label{Ed2}
\eea
The first two lines of this formula reproduce 
Eq.~(7.28) of Ref.~\cite{dG}. The constant $B$ is associated
with uniform transverse compression or dilation,
$C$ with transverse shear (Fig.~\ref{fig1}c),
and $K_3$ with bending. 

The third line contains higher-order terms.
The last one may be important
when amplitudes of longitudinal deformations exceed 
$(K_3/B'')^{1/2}$. 
The term with $B'$ provides a non-linear coupling of
transverse and longitudinal deformations,
which causes, among other effects, a breakdown of linear 
elasticity and hydrodynamics
at large scales 
(see \S\S 8.3--8.4 of Ref.~\cite{dG}). 
In the case of smectics, an analogous coupling is
provided by the cross term  coming from the expression in
square brackets in Eq.~(\ref{Ed1}). 

The terms $B$, $B'$, and $B''$ stem from the energy increase
due to the change of the cross section of the unit cell. 
A longitudinal shear $\partial u_x/\partial z =\Delta$
accompanied by uniform transverse compression
$\partial u_x/\partial x =\partial u_y/\partial y=\Delta'/2$
maintains the triangular lattice but leads 
to a change of the lattice spacing proportional to
$(1+\Delta')^{1/2}/(1+\Delta^2)^{1/4}$. 
Then Eqs.~(\ref{dE}) and (\ref{Ed2}) yield
$B=\frac{3}{2}E_{C0}$, $B'=-\frac{3}{4}E_{C0}$, 
and $B''=\frac{3}{8}E_{C0}$. 

The two other constants are associated with deformations 
that are not a simple scaling of the triangular lattice.  
We estimate them in a way analogous to that for smectics. 
Imposing a perturbation of the form 
${\bf u}(z)={\bf u}_0\cos kz$ and Fourier transforming,
we arrive at the expression
\be
  \langle E_C\rangle = 
  {\sum_{lmn}}' \,{2\pi w\rho_p^2\,
  J_n^2({\bf k}_{lm}\cdot{\bf u}_0) \over
  k_{TF}^2 + k_{lm}^2 + (nk)^2}
  \left(2J_1(k_{lm}r_N)\over k_{lm}r_N\right)^2,
\label{EC2}
\ee
where ${\bf k}_{lm}=l{\bf k}_1+m{\bf k}_2$ is a reciprocal lattice 
vector. In the absence of transverse deformation,
${\bf k}_1$ and ${\bf k}_2$ are vectors of length
$(8\pi/\sqrt{3})^{1/2}r_c^{-1}$, with an angle $\pi/3$ between them
(Fig.~\ref{fig1}c; we shall assume that ${\bf k}_2$ is directed along $Oy$).
With $u_0=0$, only the term $n=0$ survives, 
and neglecting $k_{TF}$,
we recover Eq.~(10) of Ref.~\cite{ohy} for $E_{C0}$.
The analytic formula
obtained by replacing the hexagonal unit cell
by an equivalent cylinder~\cite{rpw}, 
\be
  E_{C0} =(\pi/2)\, (\rho_p r_N)^2 w\,(\ln(1/w) -1 + w),
\label{ECrpw}
\ee
turns out to be very accurate at the values of $w$
at which the ``spaghetti" phase is expected\footnote{
The columnar
phase is favored for $w$ between a lower limit of 0.15--0.20 
and an upper limit of 0.30--0.35
(and an upper limit of $w$ for the layered phase is about
0.60--0.65)~\cite{rpw,hashimoto,ohy,wk85}.}:
it underestimates $E_{C0}$ for the hexagonal cell
by less than 0.6\% at $w<0.35$,
which is several times 
smaller than the difference due to
inclusion of a realistic screening wavenumber 
$k_{TF}\approx0.4\,r_c^{-1}$. 
Since for the columnar case the surface energy  
is $E_{\sigma}=2w\sigma/r_N$, 
one finds $r_N=(2\sigma)^{1/3}(\pi \rho_p^2
[\ln(1/w)-1+w])^{-1/3}$ in equilibrium. 

\begin{figure}[h]
\begin{center}
\leavevmode
\epsfysize=65mm
\epsfbox[135 300 475 550]{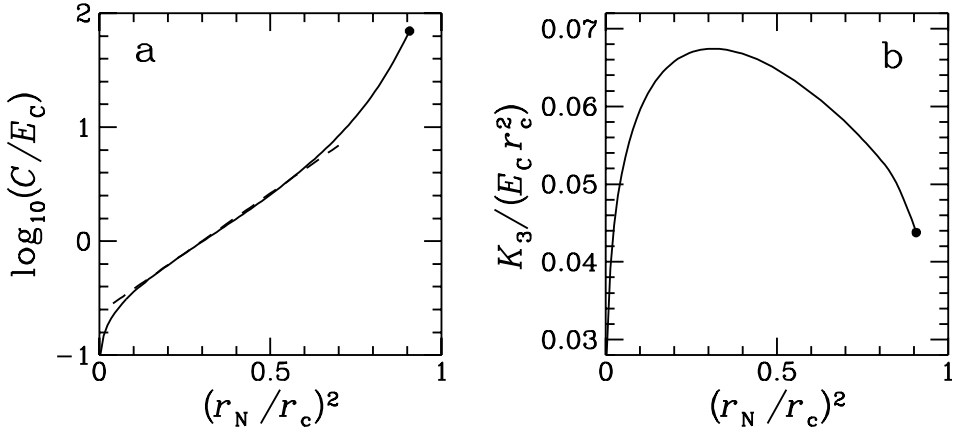}
\end{center}
\caption{Shear constant $C$ 
and bend constant $K_3$ 
of the columnar phase.}  
\label{fig:K3}
\end{figure}

A transverse shear $\partial u_x/\partial y=\Delta$
changes $k_{1y}$ by $(2\pi\sqrt{3})^{1/2}\Delta/r_c$ (Fig.~\ref{fig1}c).
Numerical summation of the series (\ref{EC2})
and identification of the term quadratic in $\Delta$
yields the elastic constant $C$, 
shown in Fig.~\ref{fig:K3}a. 
In the relevant range of $w$,
it can be approximated as 
$\log_{10}(C/E_{C0})\approx 2.1(w-0.3)$ (dashed line). 
(This estimate of $C$ is only a first approximation, 
since we have kept the cross sections of rods circular,
whereas in reality they can adjust their form 
to the shear).

Finally, using a finite ${\bf u}_0$ in Eq.~(\ref{EC2}),
performing the summation and identifying
the term proportional to $k^4 u_0^2$, we determine 
the bend constant $K_3$, which is plotted in Fig.~\ref{fig:K3}b.
In the physically relevant range of $w$, $K_3=(0.064-0.067)\,E_{C0}r_c^2$.
The term proportional to $k^4 u_0^4$
yields for $B''$ the value obtained earlier. 

The fact that matter in neutron stars near the boundary of the core has the
elastic properties of liquid crystals rather than a crystalline solid will
have important consequences for a number of aspects of neutron star behavior.
First, the maximum elastic energy that can be stored in the crust will be
reduced, and this should be taken into account in models of glitch phenomena
and starquakes that depend on the deviations of the figure of the star from
that for a fluid
(see, for example, Ref.~\cite{alpar}).  
It is not possible to make a quantitative
estimate
of the reduction of the elastic rigidity of the outer parts of the star
without a detailed model for the orientation of the liquid-crystal phases, but
since these phases are estimated to make up roughly half of the matter by mass
outside the core, one might expect the effective elastic rigidity to be
reduced by roughly a factor of 2. Since the``plate tectonics" of liquid
crystals is likely to be very different from that of crystalline solids, a
second problem where the elastic properties will be crucial is in models of
the evolution of neutron star magnetic fields that invoke such
processes~\cite{ruderman}.  Other applications are to the propagation of
elastic distortions, and to the energy of defects in the liquid-crystal
structure.

 In our discussion above we assumed that in the absence of 
deformation the spacings of rods and slabs had their equilibrium values.   
However, in reality, this may not be the case, since to alter the spacing 
requires a major rearrangement of the proton distribution.  
Just how easy it is to do this depends on how slabs and rods 
are connected to each other, and the related question of what defects 
are present.  Should the spacing be smaller than its equilibrium value, 
there will be a positive contribution to the elastic energy of the form
$(E_{\sigma}/2 - E_{C}) (\nabla_{\perp} u)^2$, 
which gives a restoring force linear in the deformation. 
Should the spacing be larger, this contribution will be negative,
which gives a tendency to rotation and leads to spontaneous deformation 
analogous to the Helfrich effect in conventional liquid crystals~\cite{dG}. 
Eventually the spacing will approach equilibrium, 
either by spontaneous reconnection of rods and slabs or by motion of defects.

To summarize, the above considerations indicate that matter near 
the boundary with the core of a neutron star
has the elastic properties of a liquid crystal, 
rather than a conventional solid.  To
distinguish this region from that further out in the star, we consider it
appropriate to refer to it as the ``mantle" 
rather than as part of the ``crust".
It is clear that the properties of neutron stars need to be reconsidered in
light of this new understanding. 

\vspace{2ex}
This work was supported in part by NSF grants AST93-15133 and AST96-18524, 
NASA grant NAGW-1583, RFBR grant 96-02-16870a,
INTAS grants 94-3834 and 96-0542, and RFBR-DFG grant 96-02-00177G.
The visit of AYP to Nordita was made possible
by Nordita's Baltic/NW Russia Fellowship programme. 
We are grateful to D.~G. Ravenhall for helpful discussions. 
During the preliminary stage of this work,
one of us (CJP) had useful discussions with 
Gregor Bergman, Kristinn Johnsen, and
Mikael Sahrling.

\end{document}